\documentclass[
]{ceurart}

\usepackage{listings}
\usepackage{todonotes}

\usepackage{xcolor}
\usepackage{listings}
\usepackage{textcomp}
\usepackage{xcolor}
\newcommand{\pleg}{{\small{\sf PROLEG}}}
\newcommand{\pl}{{\small{\sf PROLOG}}}
\newcommand{\pbl}{{\sf ProbLog}}
\newcommand{\pbleg}{{\sf ProbLEG}}

\begin{document}

\copyrightyear{2023}
\copyrightclause{Copyright for this paper by its authors.
  Use permitted under Creative Commons License Attribution 4.0
  International (CC BY 4.0).}

\conference{London'23: Workshop on Logic Programming and Legal Reasoning,
  July 09--10, 2023, London, UK}

\title{Beyond Logic Programming for Legal Reasoning}

\tnotemark[1]

\author[1]{Ha Thanh Nguyen}[%
orcid=0000-0003-2794-7010,
email=nguyenhathanh@nii.ac.jp
]
\cormark[1]
\address[1]{National Institute of Informatics (NII),
  2-1-2 Hitotsubashi, Chiyoda City, Tokyo, Japan}


\author[2]{Francesca Toni}[%
orcid=0000-0001-8194-1459,
email=f.toni@imperial.ac.uk,
]
\address[2]{Imperial College London, Exhibition Rd, South Kensington, London SW7 2BX, United Kingdom}

\author[3]{Kostas Stathis}[%
orcid=0000-0002-9946-4037,
email=Kostas.Stathis@rhul.ac.uk,
]
\address[3]{Royal Holloway University of London, Egham Hill, Egham TW20 0EX, United Kingdom}

\author[1]{Ken Satoh}[%
orcid=0000-0002-9309-4602,
email=ksatoh@nii.ac.jp,
]
\cortext[1]{Corresponding author.}

\begin{abstract}
Logic programming has long being advocated for legal reasoning, and several approaches have been put forward relying upon explicit representation of the law in logic programming terms. In this position paper we 
focus on the \pleg\  logic-programming-based framework for formalizing and reasoning with Japanese presupposed ultimate fact theory. 
Specifically, we 
examine challenges and opportunities in leveraging deep learning techniques for improving legal reasoning using \pleg, 
identifying 
four distinct options 
ranging from enhancing fact extraction using deep learning to end-to-end solutions for reasoning with textual legal descriptions. We assess  advantages and limitations of each option, considering their technical feasibility, interpretability, and alignment with the needs of legal practitioners and decision-makers. We believe that our analysis can serve as a guideline for developers aiming to build effective decision-support systems for the legal domain, while fostering a deeper understanding of challenges and potential advancements 
by neuro-symbolic approaches in legal applications.
\end{abstract}

\begin{keywords}
  PROLEG \sep
  Deep Learning \sep
  ProbLog \sep
  Prolog \sep
  Neurosymbolic \sep
  Legal Reasoning
\end{keywords}

\maketitle

\section{Introduction}

Legal reasoning is a complex process that involves evaluating and applying principles, rules, and regulations from various sources, such as legislation, case law, and general legal principles. The advancement of AI, particularly in terms of deep learning and natural language understanding, has created new opportunities for developing systems that can aid legal practitioners in navigating this complex landscape.

In this paper, we examine the challenges and opportunities in leveraging deep learning techniques for improving legal reasoning using the \pleg\ system \cite{satoh2010proleg}, a logic-programming-based framework for formalizing and reasoning with the Japanese presupposed Ultimate Fact theory (\emph{JUF theory}, in short, from \emph{Youken-jijisturon}, in Japanese). The JUF theory is used for decision-making by judges under incomplete information. 

\pleg\ accommodates a representation of 
the JUF theory in logic programming terms, 
reflecting lawyers' reasoning by drawing upon the idea of ``openness'' proposed 
for the JUF theory.
However, the deployment of \pleg\ to reason about particular input cases requires the addition of suitable facts, relating to the cases, to the system.  

In this position paper we explore various options for integrating deep learning techniques into the legal reasoning process envisaged by \pleg. We discuss these options in the context of extracting and leveraging facts from natural language texts, ranging from enhancing fact extraction to end-to-end solutions for reasoning with textual legal descriptions.
We provide toy illustrations of the various options.

Our analysis is targeted at developers working on decision-support systems for legal practitioners, and our proposed options are designed to address the challenges and limitations observed in related works, such as understanding and applying prescriptive rules specified in natural language, providing better structural insights, and effectively representing and sharing legal knowledge using well-defined logical languages.

Throughout the paper, we assess the advantages and limitations of each option, considering their technical feasibility, interpretability and alignment with the needs of legal practitioners such as judges, jurors, and lawyers involved in decision-making based on legal information. By combining deep learning with symbolic reasoning methods, we aim to provide developers with insights and solutions for building effective and comprehensible decision-support systems in the legal domain.








\section{Related Work}

Natural Language Processing (NLP) and Natural Language Understanding (NLU) tasks applied to law have proliferated in the recent years, with several works in the literature focusing on the challenge of legal reasoning and decision-making using structured, logical representations derived from texts. This work relates to several areas of active research. 

Holzenberger et al. \cite{holzenberger2020dataset} presents a dataset and legal-domain text corpus to investigate the performance of NLU approaches on statutory reasoning. The study compares the results of machine reading models with a hand-constructed Prolog-based system, highlighting the challenges facing statutory reasoning moving forward. This work demonstrates the importance of understanding and applying prescriptive rules specified in natural language for legal reasoning. Further work by Holzenberger and Van Durme~ \cite{holzenberger2021factoring} decomposes statutory reasoning into four types of NLU challenge problems by introducing concepts and structures found in Prolog programs. 
Their results show that models for statutory reasoning benefit from the additional structure, leading to improved performance over prior baselines and finer-grained model diagnostics.

Palmirani et al. \cite{palmirani2011legalruleml} extends the RuleML language \cite{boley2010ruleml} to create LegalRuleML, enabling the effective sharing and exchange of legal knowledge between documents, business rules, and software applications. This work focuses on detecting, modeling, and expressively representing legal knowledge to support legal reasoning and its application in the business rule domain. In a similar context, Schneider et al. \cite{schneider2022lynx} describe the EU-funded project Lynx, which aims to create a Legal Knowledge Graph (LKG) for the semantic processing, analysis, and enrichment of legal documents. The article discusses the use cases, platforms, and semantic analysis services that operate on the documents for more effective legal information management.

Early work by Nakamura et al.~\cite{transl-legal} proposes a framework for translating legal sentences into logical form, and describes the implementation of such an experimental system. In a similar vein, Lagos et al.~\cite{lagosetal} present an event extraction mechanism  based on NLP techniques to extract the use of entity related information corresponding to the relations among the parties of a case in the form of events. More recently, the work by Gaur et al. \cite{gaur2015translating} address the translation of legal text into formal representations using the NL2KR system. This system translates legal text into various formal representations, enabling the use of existing logical reasoning approaches on legal text in English. By allowing reasoning with text, this work bridges the gap between natural language processing and existing logical reasoning frameworks designed for legal information.

In the remainder, we will explore several options for improving the performance and applicability of the \pleg\ legal reasoning support system \cite{satoh2010proleg}. Each option proposed is based on the related work and methods discussed above, aiming to enhance the reasoning capabilities of \pleg\ by incorporating deep learning techniques for extracting and leveraging facts from natural language texts.
These options attempt to address different challenges and limitations observed in previous works, such as the difficulty in understanding and applying prescriptive rules specified in natural language~\cite{holzenberger2020dataset}, the need for additional structural insights provided by Prolog-based systems~\cite{holzenberger2021factoring}, and the importance of effectively representing and sharing legal knowledge using well-defined logical languages~\cite{palmirani2011legalruleml}.


\section{Options}

We focus here on the following setting: laws have been mapped onto \pleg; there are text descriptions of the ``facts'' to which the laws need to be applied; we want to ``emulate'' reasoning with the laws and facts by using \pleg\ and a logical description of the ``facts''.\footnote{This setting is in line with the goals of a \emph{judge} or \emph{juror} who needs to decide on the outcome of a case.  Of course, there are other settings of interest. For example, we could be wanting to generate the ``facts'' so that the outcome of the application of the  law is ``driven'' by the lawyer presenting the ``facts''.
In this alternative setting, we may want to look for most ``beneficial facts''. This latter setting is in line with the operation of defence or plaintiff lawyers. }

\subsection{Option 1: \pleg\ + Fact Extraction by Deep Learning}
This is the approach envisaged with \pleg, whereby the facts are standard \pl\ facts and are to be used in combination with \pleg\ using its standard inference mechanism. 
The fact extraction is carried out by fine-tuning a language model.

Fact extraction requires 
carrying out several sub-tasks, given an input text (the query) describing the facts in natural language:

\begin{itemize}
    \item Identifying which (article of) law matches the query;
    \item Identifying which bits of the  (article of) law matches the query;
    \item Populate the logical facts underpinning the law with the information from the query matching the law (see \cite{nguyen2022multi,may2023improving}).
\end{itemize}
\subsection{Option 2: \pbleg: \pleg\ + Probabilistic Fact Extraction by Deep Learning}
This is an alternative approach,  whereby the facts are probabilistic as in \pbl~\cite{ProbLog15} and are to be used in combination with \pleg\ using \pbl's inference mechanism. 
The fact extraction needs to be again carried out externally to \pleg\ and \pbl, e.g. by fine-tuning a language model \cite{vuong2022sm,nguyen2022attentive,nguyen2022transformer,shao2020bert}.

\subsection{Option 3: Deep ProbLEG}

The first two approaches are neuro-symbolic in a loose sense (in that the deep learning and reasoning components interface but are kept separate).
A fundamentally different approach integrates reasoning and fact extraction tightly in the spirit of Deep \pbl~\cite{ManhaeveDKDR21}. This approach would require injecting \pleg\ into a deep architecture of the kind explored in \cite{ManhaeveDKDR21}, whereby reasoning a-la-\pbl\ follows probabilistic fact extraction within a cohesive whole.

\subsection{Option 4: End-to-End Deep Learning}

This final approach is aimed at serving as a baseline and is expected to be less 
interpretable than the other three, based on fine-tuning language models,
such as domain-oriented models \cite{nguyen2020jnlp,chalkidis2020legal} or robust domain-independent models \cite{raffel2020exploring,brown2020language,chowdhery2022palm},
to reason directly with the textual representation of the legislation with the textual representation of the facts. This approach would not rely on reasoning with \pleg\ directly and will be  non-interpretable
.

Methodologically, we could see the legal reasoning that \pleg\ is carrying out as a form of \emph{soft theorem proving}  \cite{softtheorem}, possibly fine-tuning the adversarial method in \cite{alex}.

\section{A toy illustration}
\label{sec:toy}

Here we focus on illustrating the first two options with a simple toy example with standard \pleg\ structure but made-up laws. Our aim is to show what the two options may amount to. 

\noindent A first law may be represented in \pleg\ by:\\

\vspace{-.8cm}

\begin{lstlisting}
right(B,S)<=
 cond(B,S).

cond(B,S) <=
 sub_con(B,S). 
\end{lstlisting}
\noindent $sub\_cond$ may amount to `being friends'.

\begin{lstlisting}
exception(sub_con(B,S),ex(B,S))
ex(B,S) <=
  ex_con(B,S) 
\end{lstlisting}
$sub\_cond$ may amount to `knowing each other for a long time'.

\vspace{1cm}
\noindent A second law may be represented simply as:\\
\begin{lstlisting}
other_right(B,S) <= smtg(B,S)
\end{lstlisting}
$ex\_cond$ may amount to `working together'

\noindent Facts are about 
$sub\_con(B,S)$
and $ex\_con(B,S)$,
as well as
$smtg(B,S)$.
\\

\noindent
Suppose the textual description of  three cases of interest is as follows:

\begin{description}
\item[case 1]  Thanh and Kostas met recently but are good friends already (\emph{thus they satisfy the sub-conditions for the first law}); 
\item[case 2] Ken and Francesca 
have collaborated for a while (\emph{do they  satisfy the conditions of the second law? the exception to the first law?}) and are in good terms (\emph{do they satisfy the conditions of the first law?}); 
\item[case 3]  Thanh is working under Ken's supervision (\emph{do they trigger the second law?}) and they are in good terms (\emph{do they satisfy the sub-conditions of the first law?}).
\end{description}

Here, case 1 most definitely matches the first law, and case 3 probably matches the second law (but may also be deemed somewhat to match the first law); it is very unclear though whether case 2 matches the first law (with exception) or the second law. Also,  after deciding which law case 2 matches, there is uncertainty in extracting the facts.

\paragraph{Mapping \pleg\ 
into \pl.} 
We focus here on the  object-level translation of the above \pleg\ program~\cite{satoh2010proleg}, as follows:\\

\begin{lstlisting}
right(B, S) :- cond(B, S).
cond(B, S) :- sub_con(B, S), not ex_con(B, S).
other_right(B, S) :- smtg(B, S).

sub_con(thanh, kostas).
sub_con(ken, fran).

ex_con(ken, fran).
smtg(thanh, ken).
\end{lstlisting}

\noindent This translation makes use of transforming exception clauses in \pleg\ as negation by failure in the condition of the rules\footnote{This is compatible with the spirit of the standard translation methods for \pleg\, already described in  http://research.nii.ac.jp/~ksatoh/juris-informatics-papers/jurix2009-ksatoh.pdf}. An alternative translation that we may want to consider is given in Appendix~\ref{app:other translation}.

\paragraph{Option 1}
Deep learning learns ``crisp'' logical facts, e.g. they may be:\\

\begin{lstlisting}
sub_con(thanh,kostas). %(first_case)

sub_con(ken,fran). %(and)
ex_con(ken,fran). %(second_case)

smtg(thanh,ken). %(third_case)

\end{lstlisting} 

\paragraph{Option 2} 
Deep learning learns probabilities. What are these probabilities on? The facts or the applicability of the laws? This makes a difference if $(cond(B,S) \cup ex\_con(B,S)) \cap smtg(B,S)\neq \emptyset$.
Also, even if $(cond(B,S) \cup ex\_con(B,S)) \cap smtg(B,S)= \emptyset$, the probability of matching a law may play a role in determining the probability of the facts holding.

\paragraph{Determining the probability of facts in isolation.}
This may result in  something like\\

\begin{lstlisting}
0.99::sub_con(thanh, kostas).

0.51::sub_con(ken, fran).

0.73::ex_con(ken, fran).

0.49::smtg(ken, fran).

0.51::sub_con(thanh, ken).

0.65::smtg(thanh, ken).
\end{lstlisting}

\paragraph{Determining the probability of facts while taking into account the relevance of laws.}
This means that the probability of a fact may be a function of two probabilities: that the case matches a law and that phrases in the case description fit atomic templates. For example,
we may compute the following probabilities that the cases match the laws:\footnote{We may need to assume that the two laws are independent.}\\

\begin{itemize}
    \item Case 1 matches law 1 with probability 1;
    \item Case 2 matches law 1 with probability 0.7 and law 2 with probability 0.3;
    \item Case 3 matches law 2 with probability 0.6 and law 1 with probability 0.4.
\end{itemize}

Then, for each case (when the law applicability is probable according to some treashold), we determine the probability of 
the atomic templates $sub\_cond(\cdot,\cdot)$,
$ex\_cond(\cdot,\cdot)$, and
$smtg(\cdot,\cdot)$
being instantiatable on the textual description of the cases.

Suppose that case 1 matches $sub\_cond(\cdot,\cdot)$ with probability 0.99. Then, the resulting probabilistic fact is\\

\begin{lstlisting}
(1 * 0.99) = 0.99::sub_con(thanh, kostas).
\end{lstlisting}

Suppose that case 3 
matches $smtg(\cdot,\cdot)$
with probability 0.9. Then, the resulting probabilistic fact is\\

\begin{lstlisting}
(0.6 * 0.9) = 0.54::sub_con(ken, fran).
\end{lstlisting}

\paragraph{Keeping the relevance of the laws and the matching of the atomic templates separate.}
This amount to treating the applicability of laws ``abductively'', and may lead (in our example) to:\\

\begin{lstlisting}
right(B, S) :- cond(B, S), applicable_1(B, S).

cond(B, S) :- sub_con(B, S), not ex_con(B, S).

other_right(B, S) :- smtg(B, S), applicable_2(B, S).
\end{lstlisting}

\noindent Then the fact:

\begin{lstlisting}
0.95::sub_con(thanh, kostas).
0.55::sub_con(ken, fran).

0.78::ex_con(ken, fran).
0.42::smtg(ken, fran).

0.50::sub_con(thanh, ken).
0.69::smtg(thanh, ken).

0.99::applicable_1(thanh, kostas).

0.51::applicable_1(ken, fran).
0.49::applicable_2(ken, fran).

0.71::applicable_1(thanh, kenn).
0.81::applicable_2(thanh, kenn).
\end{lstlisting}

\section{Discussions and Conclusion}
We have explored challenges and opportunities
of combining deep learning techniques and the PROLEG system to improve legal reasoning. Our analysis has identified four distinct options ranging from enhancing fact extraction using deep learning to end-to-end solutions
for reasoning with textual legal descriptions. Throughout our analysis, we have taken the standpoint of developers rather than legal practitioners. In other words, the systems we envision could be used as the backend of decision-support systems, potentially by enforcing certain threshold conditions to facilitate decision-making and reasoning in legal contexts.

Although we primarily focused on a "judge/juror" setting for developers, some of the analyses and methodologies we propose could be adapted to a "lawyer" setting, as discussed earlier. In this context, our systems could be used to support the selection and presentation of facts that lead to a desired conclusion (or a high probability of it) and aid in evidence retrieval to build strong legal arguments and cases.

By integrating deep learning with symbolic reasoning methods, we can develop more robust and flexible legal reasoning systems that can handle the challenges posed by natural language ambiguity and rigid logical representations.

Some additional benefits of the proposed approaches include:

\begin{itemize}
    \item A better understanding of the strengths and weaknesses of different arguments and the underlying evidence.
    \item Effective identification of crucial evidence and the areas where such evidence can be found to overturn a legal decision if necessary.
    \item Enhanced capability to debug legal decisions and regulations, allowing for more efficient evaluation and adjustments in legal frameworks.
\end{itemize}

In conclusion, by leveraging the advancements in deep learning and natural language understanding, the proposed options aim to provide developers with practical tools and methods for building decision-support systems that aid various legal practitioners, including judges, jurors, and lawyers, in navigating the complex landscape of legal reasoning and decision-making. Moving forward, we plan to  delve deeper into the four options, providing a comprehensive examination and evaluation of each method to further enhance our understanding of their potential impact and applications in the legal domain. 





\begin{acknowledgments}
This work was supported by JSPS KAKENHI Grant Number JP22H00543 and JST, AIP Trilateral AI Research Grant Number JPMJCR20G4.
  Francesca Toni and Kostas Stathis would like to thank the National Institute of Informatics, Tokyo, Japan, for supporting their visit to Japan that made this work possible. Francesca Toni also acknowledges support from the European Research Council (ERC) under the European Union’s Horizon 2020 research and innovation programme (grant agreement No.101020934, ADIX), as well as support from J.P. Morgan and the Royal Academy of Engineering, UK, under the Research Chairs and Senior Research Fellowships scheme. 
\end{acknowledgments}


\bibliography{sample-ceur}

\appendix



\section{Alternative \pleg\ translations}
\label{app:other translation}

A somewhat more sophisticated object-level translation of the  \pleg\ program in Section~\ref{sec:toy} could be as follows:\\

\begin{lstlisting}
right(B, S) :- cond(B, S).

cond(B, S) :- sub_con(B, S), not exempt_sub_con(B, S).

exempt_sub_con(B, S) :-
  ex_con(B, S), not exempt_ex_con(B, S).

other_right(B, S) :- smtg(B, S).

sub_con(thanh, kostas).

sub_con(ken, fran).

ex_con(ken, fran).

smtg(thanh, ken).
\end{lstlisting}
\noindent The latter translation method is more methodological and has hooks for exceptions of exceptions.



\end{document}